# Twist as a Mechanical Switch for Reconfigurable Stacking in *h*-BN

Gautham Vijayan[1†], Zhaoheng Zhang[2†], Kenji Watanabe[3], Takashi Taniguchi[3], Michael Urbakh[4], Oded Hod[4], Xiang Gao[2*], Elad Koren[1*]

[1]*Nanoscale Electronic Materials and Devices Laboratory, Faculty of Materials Science and Engineering, Technion - Israel Institute of Technology, Haifa, 3200003, Israel.*

[2]*Department of Modern Mechanics, School of Engineering Science, University of Science and Technology of China, Hefei, Anhui 230026, China*

[3]*International Center for Materials Nanoarchitectonics, National Institute for Materials Science, Tsukuba, Japan*

[4]*Department of Physical Chemistry, School of Chemistry, The Raymond and Beverly Sackler Faculty of Exact Sciences and The Sackler Center for Computational Molecular and Materials Science, Tel Aviv University, Tel Aviv 6997801, Israel*

[†]These authors contribute equally to this work.

[*]Corresponding authors. Email: xianggao@ustc.edu.cn, eladk@technion.ac.il

## Abstract

Twistronics of layered materials has emerged as a highly active field due to its profound implications for quantum electronics and materials engineering. However, the controllable manipulation of interlayer stacking remains a significant experimental challenge. Here, the homogeneous contact between hexagonal boron nitride layers is shown to be reproducibly switched between two distinct stable stacking configurations via an externally applied torque. Combining experiments and computational modelling, we identify the stacking order of these stable states as the commensurate AA′ and AB modes. These two states are associated with different rotational torque maxima, exhibiting an asymmetry ratio of ~0.7, and distinct dynamics as a function of the twist angle. Moreover, the peak torque values scale linearly with contact area, highlighting the dominant role of edge elasticity in the twisting process. Given that the AA′ and AB stacking modes correspond to different out-of-plane electric polarization states, our findings offer a pathway for reconfigurable nano- and micro-electromechanical devices.

## Introduction

The ability to reconfigure interfaces on demand offers a powerful pathway to emergent functionality in two-dimensional (2D) materials. Interfacial landscapes can be dynamically redefined through mechanically driven twisting[1,2,3], thermal annealing[4,5], electric-field switching[6] and local strain[7]. Distinct interfacial configurations can also be accessed through lateral sliding, which formulates the field of slidetronics, where changes in stacking registry directly modulate the interfacial polarization[8,9]. Similarly, twistronics involves a relative twist between atomically thin crystals that reshapes the interface and generates moiré superlattices with angle-dependent properties[10–13]. In graphite and its heterostructures, such angular tuning has revealed a characteristic 60° periodicity in interfacial behaviors, as observed in friction[14–18], torsion[19–21] and electrical transport[22,23]. Yet, despite the richness of the moiré landscape, graphite effectively cycles through equivalent commensurate registries, limiting the possibility of switching reversibly between distinct stacking configurations.

In contrast, hexagonal boron nitride (*h*-BN) offers a fundamentally different platform. Owing to its threefold symmetry and chemically inequivalent atomic constituents, continuous rotation connects multiple distinct commensurate stackings, separated by intervening moiré states. This creates a dynamically reconfigurable interface that can be driven from one commensurate state to a moiré state, then into a second commensurate state, and eventually back to the original registry through continued rotation. Such behavior is particularly compelling in layered systems, as it enables rotational control over multiple functional states of distinct polarization[6,8,9], charge transport[22–26], and electromechanical responses[27,28], within a single material platform.

Here, we demonstrate that twist-induced configuration switching in *h*-BN is reversible, where each commensurate state exhibits a quantitatively different rotational torque threshold. By combining experiment with theory, we identify the non-polar AA′ and the polar AB stacking modes as the stable configurations along the twist path and decipher the microscopic origin of their different torsional responses. We further find that the peak torque values scale linearly with contact area for both commensurate configurations, indicating the dominant role of interfacial elasticity at the contact edges.

**Experimental setup**

High-quality $h$-BN flakes, with thicknesses of approximately 250 nm, were mechanically exfoliated onto $SiO_2$/Si wafer and used as the substrate for all experiments. Circular metal contacts (with radii of 200, 300, 400, and 500 nm) with a lever arm, comprising 5 nm Cr, 25 nm Ni and 25 nm Au, were patterned by electron-beam lithography. Subsequent reactive ion etching using $SF_6$ defined pillar-like structures with heights of approximately 100 nm. These mesas served as the platform for all torsional measurements described herein. Full fabrication details are provided in the Supplementary Information Section (SI Sec.) 1.

The experiments were conducted using an atomic force microscope (AFM, Bruker Dimension Icon) housed inside a nitrogen filled glovebox with $H_2O$ and $O_2$ levels maintained below 1 ppm. Torsional measurements on $h$-BN were performed using a bearing-like architecture, in which the $h$-BN mesa was mechanically actuated with an AFM cantilever. This configuration enabled continuous rotation of the top $h$-BN mesa relative to the fixed bottom mesa, completing a full 360° rotation of the circular interfacial contact, as schematically illustrated in Fig. 1a. An AFM topography image of a few representative structures following 180° rotation is shown in Fig. 1b, where the corresponding height profile extracted along the red dashed line is presented in Fig. 1c. The height analysis indicates that the rotating interface forms approximately halfway along the pillar height, spatially separated from both the metal contact and the bottom surface, confirming that the interface is defined exclusively by $h$-BN layers. During rotation, the lateral deflection of the cantilever and the Y-axis piezo displacement along the circular trajectory were recorded using an oscilloscope (Keysight DSOS054A) and used to reconstruct the force-angle profiles. The lateral force constant of the AFM cantilever was calibrated using the reported adhesion energy of $h$-BN[5,19,29–31] (see SI Sec. 2), and the Y-axis piezo displacement was subsequently converted into angular displacement[19,22] (see SI Sec. 3).

During actuation, the contact angle between the cantilever and the lever arm evolved continuously, such that the force recorded by the AFM cantilever corresponded only to the horizontal ($F_x$) projection of the actual force ($F_{\mathrm{real}}$) required to generate torque at the circular interface[19]. The measured lateral force was therefore intrinsically geometry dependent: being maximal when the cantilever is parallel to the lever arm, and vanishing when the cantilever is perpendicular to it, following a cosine projection at intermediate orientations. A detailed description of the cantilever-sample interaction is provided in SI Sec. 4.

## Results and discussion

A representative force-angle profile over a full rotation is shown in Fig. 1d. Pronounced force maxima emerge at rotation angles corresponding to multiples of 60°, with the first peak appearing at 60°, followed by a larger peak at 120°. No significant peak is observed at 180°, consistent with the cantilever being vertically aligned with respect to the lever arm at this position. Larger and smaller peaks reappear at 240° and 300°, respectively. The reversal in the sign of the force peaks arises from the change in the direction of lateral cantilever deflection over the two semi-circular segments of the trajectory: one half-cycle produces positive deflection (left-to-right motion), whereas the subsequent half-cycle produces negative deflection (right-to-left motion).

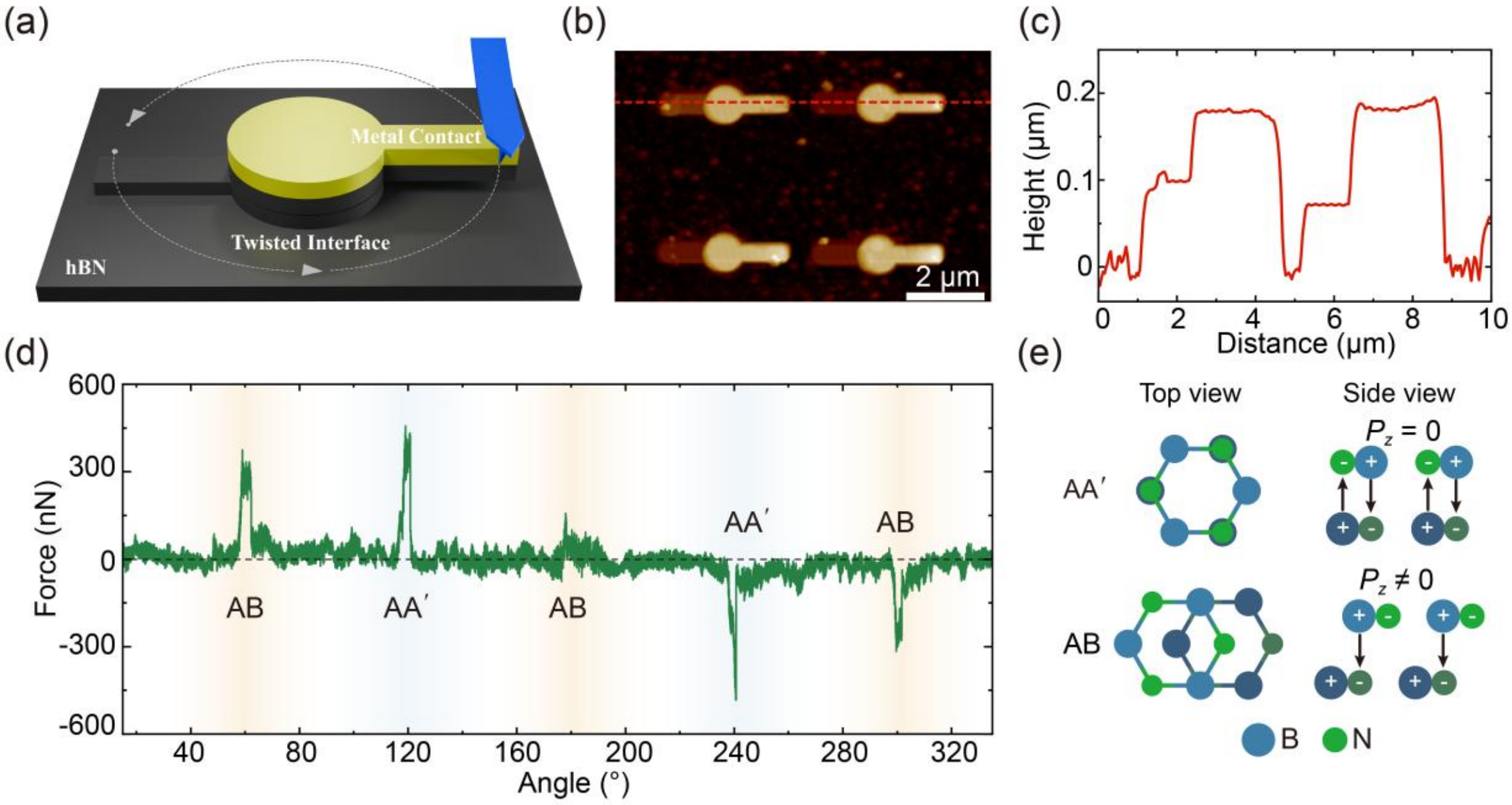

**Figure 1.** Mechanical actuation and force response of rotationally reconfigurable *h*-BN interfaces. (a) Schematic of the 360° mechanical actuation device, where the top mesa continuously rotates relative to the bottom mesa around the circular pilar centeral axis. (b) AFM topography image of four test structures following 180° rotation. (c) Height profile extracted along the red dashed line in panel b. (d) Force-angle profile recorded during mechanical actuation (for circular radius of 500 nm). The dashed line, marking zero force, serves as a guide to the eye. A full (360° range) force-angle curve including the initial pillar-breaking stage is presented in Fig. S4b. (e) Top and side view schematics of the commensurate atomic stacking configurations corresponding to the force maxima in panel d. Light and dark colored circles represnts the top and bottom layers, respectively.

These measurements reveal that the rotation does not proceed through a uniform torque landscape, but instead through an alternating sequence of lower- and higher-torque states introduced by the threefold symmetry of the $h$-BN interface. Our theoretical calculations discussed ahead assign the alternating peak magnitudes to the commensurate non-polar AA′ and metastable polar AB stacking configurations[32,33] (see Fig. 1e) with the former mediated by the nucleation of metastable AB′ stacking.

To quantify the torque ($T$) required to toggle the system between these two commensurate states we extracted the peak lateral forces ($F_{\text{max}}$) at twist angles of 60° (AB) and 120° (AA′) and used the relation $T = |\boldsymbol{T}| = |\boldsymbol{F}_{\text{max}} \times \boldsymbol{L}|$, where $L$ is the radius of the circular actuation trajectory (Fig. 2a). As mentioned above, an angular correction was applied to extract the tangential force from the lateral force component measured by the cantilever, accounting for the sample orientation during rotation (see SI Secs. 4-5).

Similar measurements were performed across multiple pillars of different contact areas, with radii ranging from 200 to 500 nm, allowing to establish the scaling of the torque maxima with interfacial area. Intriguingly, Figure 2c reveals a linear dependence of the measured torque maxima with respect to the contact area ($T \propto A$) for both AA′ and AB stacking modes over the investigated radius range. Notably, the scaling deviates qualitatively from rigid model prediction of $T \propto A^{1.5}$ (SI Sec. 6), suggesting the dominant role of interfacial elasticity. From the linear fittings, the extracted torque density (torsional energy) is 0.88 ± 0.03 nN·nm/nm$^2$ for the AB configuration at 60° and 1.28 ± 0.03 nN·nm/nm$^2$ for the AA′ configuration at 120°, yielding an asymmetry ratio $T_{\text{AB}}/T_{\text{AA}'} \approx 0.7$. This observation is rather counterintuitive considering the comparable stacking energies of the two configurations[32] and stands in contrast to the consistently symmetric torque profiles measured for graphite/graphite and $MoS_2$/$MoS_2$ interfaces[19].

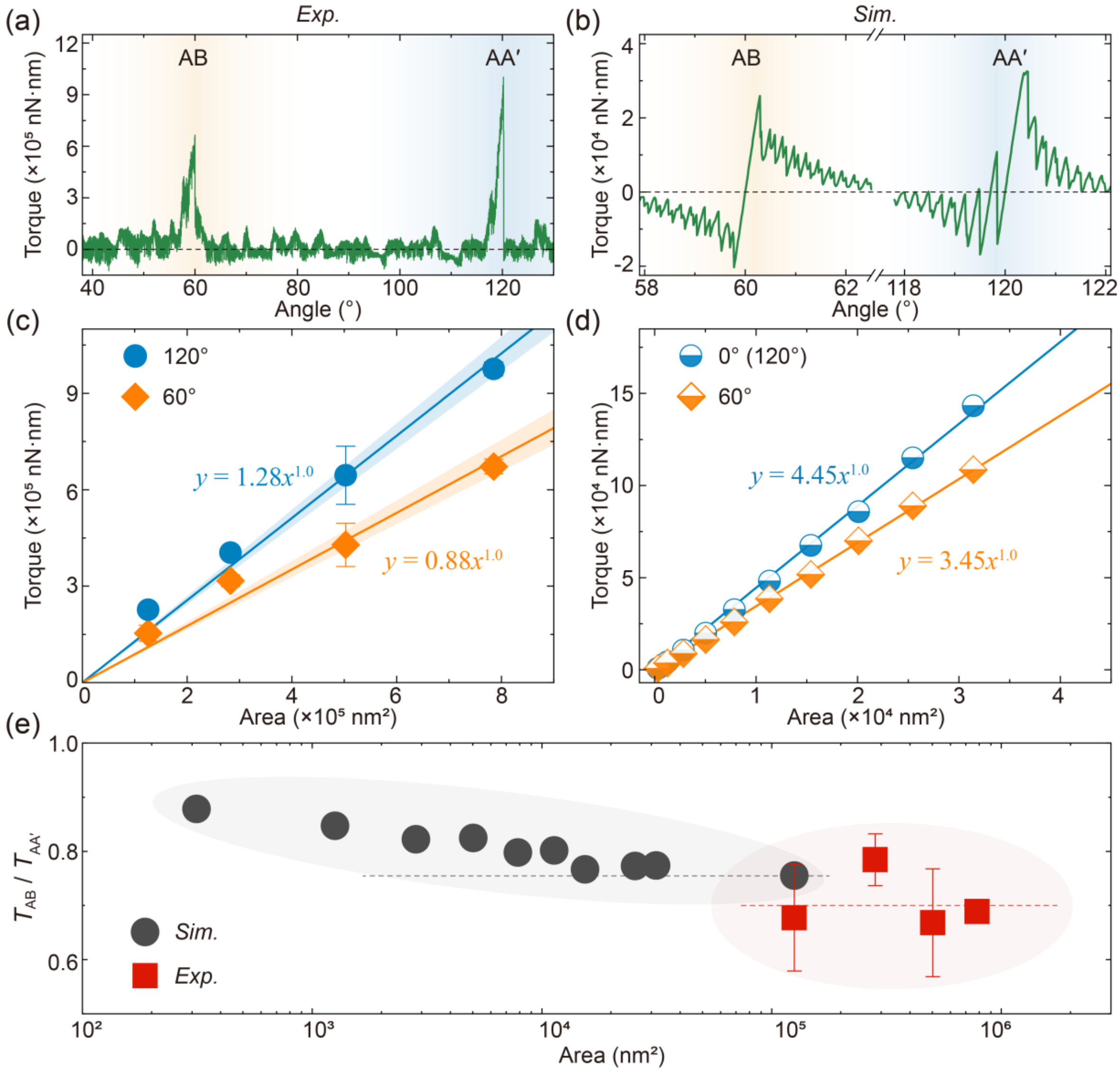

**Figure 2.** Experimental and MD simulation results for the torque-angle profiles and contact area scaling of interfacial torque in $h$-BN. (a) Experimental torque-angle profile with $R = 500$ nm showing individual force maxima associated with the AA′ and AB stacking configurations after angle correction. (b) Simulated torque-angle profile for $R = 50$ nm. (c) Experimental measurements of the contact-area scaling of the peak torque at twist angles of 60° and 120°, corresponding to the AB and AA′ stacking configurations, respectively. The shaded bands denote the 95% confidence intervals of the linear fits. Error bars were obtained from four individual measurements per point. (d) MD simulation results of the peak torque scaling for the AB and AA′ configurations. (e) The summarized torque ratio between AB and AA′ stacking, $T_{\mathrm{AB}}/T_{\mathrm{AA'}}$, for different contact area from both experiements (red squares) and MD simulations (black circles).

To reveal the microscopic physical mechanisms underlying the torsional asymmetry and peak torque scaling with contact area observed in the experiments, fully atomistic quasi-static simulations were performed using the LAMMPS package[34]. To mimic the experimental actuation via a lever arm, a three-layer model was employed, consisting of a flexible extended $h$-BN substrate, a flexible finite circular $h$-BN flake, and a rigid driving circular ring with a width of 1 nm replicated from the flake, as shown in Fig. 3a. The replicated atoms in the driving ring were coupled to their original counterparts in the flake by harmonic springs in the lateral directions, with an effective stiffness that approximates the shear modulus of the interface of two contacting graphene layers. Same lateral harmonic constraints were also applied to the substrate atoms. Moreover, to mimic the effect of multilayer stacks, harmonic constraints with an effective stiffness matching the normal modulus of bilayer graphene (see SI section 7 for further details regarding the effective springs) were applied to the vertical degrees of freedom of both the substrate and flake atoms. The intra- and interlayer interactions were described via the Tersoff[35] and ILP[36–38] potentials, respectively. In the quasistatic simulations, torsional loading was applied via incremental rigid rotation of the driving ring. Following each angular step, the system energy was relaxed using a combination of the conjugate gradient and FIRE algorithms[39,40]. Further simulations details are provided in SI Sec. 7.

By considering the possible rotation scenarios consistent with the observed threefold symmetry, we infer that the experimentally realized rotational pathway most likely evolves from AA′ stacking at 0° to AB stacking near 60°, then back to AA′ stacking at 120° (denoted as AA′→AB→AA′, see SI Sec. 8). As shown in Fig. 2b, the simulated torque-angle profiles for a 50-nm-radius circular $h$-BN flake under this path well reproduces the experimentally observed asymmetric periodic torque characteristics. The primary torque peak near 0° (120°) (most stable AA′ stacking) reaches approximately 32.58 μN·nm, while the secondary peak near 60° (stable AB stacking) gives a lower value of 25.91 μN·nm. We note that the threefold symmetry is preserved only for rotation centers at high symmetry sites, while absent for rotation around low-symmetry sites (SI Sec. 9).

We further compare the contact-area scaling of the interfacial torque near 0° (AA′) and 60° (AB). As shown in Fig. 2d, the simulated results exhibit a linear dependence of peak torque with contact area ($T \propto A$), consistent with the experimental results. The extracted effective torsional energies are 4.45 and 3.45 nN·nm/nm$^2$ for the 0° and 60° states, respectively, converging to a torque ratio $T_{\mathrm{AB}}/T_{\mathrm{AA'}} \approx 0.76$ for large radii (Fig. 2e). Although the absolute values of the simulated torque densities are 3-4 times larger than the experimental values (Fig.

2c-d), their asymmetry ratio shows good consistency with the experimental measurements (Fig. 2e).

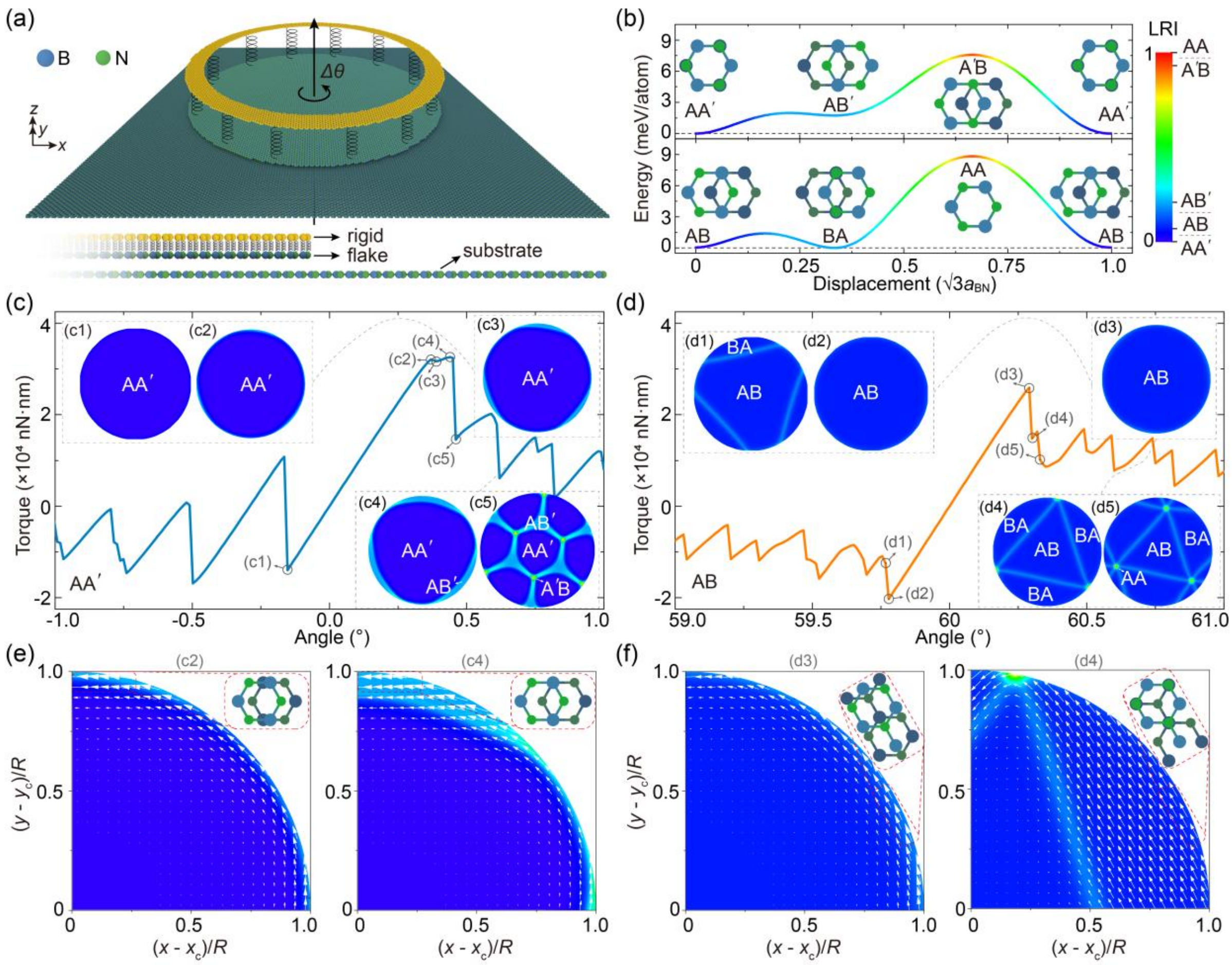

**Figure 3.** Atomistic origin of the asymmetric periodic torque behavior. (a) Schematic of the simulation setup showing the rigid driving ring harmonically coupled to the flexible circular *h*-BN flake. (b) Sliding energy curves for anti- and parallel stacking configurations calculated by rigid shift of *h*-BN layers at the equilibrium interlayer distance of each point. The curve colors correspond to the LRI value, as depicted by the color bar on the right. (c)-(d) Zoom-in views of torque-angle profiles at around (c) 0° at AA′ and (d) 60° at AB obtained from flexible quasistatic MD simulation for the circular flake of $R = 50$ nm corresponding to panel (b) in Fig. 2. The insets depict the marked instantaneous stacking configurations manifested by LRI. (e)-(f) Superimposed LRI maps with atomic displacement fields for the upper-right quarter of the flake corresponding to the selected snapshots in panels (c) and (d). Atomic displacements are calculated relative to their equilibrium states. Insets show the corresponding local stacking configurations in the dashed regions. The arrow size depicts the magnitude of local displacement.

To unveil the origin of the asymmetry in peak torque, we analyze the torque-angle profiles together with local registry index (LRI)[41–43] mapping the structural evolution across the sheared interface (LRI varies oppositely with stacking stability (Fig. 3b)). The zoom-in views of the torque-angle profiles in Fig. 3c-d demonstrate a clear difference in the yielding modes between the two commensurate states. Near 0° (AA′), torque first grows linearly, then yields with a weaker growth rate, reaches the peak, and drops slightly until a sudden slip occurs, analogous to ductile fracture. Correspondingly, the LRI mapping (Supplementary Movie 1) reveals that during this process, the interface initially stays in AA′ stacking (c2), then at the yielding point, AB′ stacking begins to enter from the rim (c3) and propagates inward until rotation slip occurs (c4). Figure 3e demonstrates that the yielding process involves progressive lateral slip from AA′ to AB′ until the global rotation slip occurs. In contrast, near 60° (AB), torque grows linearly and yields abruptly at the peak point, similar to brittle fracture. The LRI mapping (Supplementary Movie 2) shows that the interface sticks to AB stacking (d2→d3) and then directly slips with AA and BA stacking regions entering from the edge (d4). The transition from AB to BA undergoes mainly by local lateral slip, as indicated by the atomic displacement field in Fig. 3f. The combined analysis suggests that the ductile-like yielding near 0° involves a gradual phase boundary propagation, while the brittle-like yielding near 60° involves an abrupt rotational slip from uniform to mixed stacking configurations. This yielding contrast is even more pronounced in the torque-angle profiles for larger flakes (SI Sec. 10 and (Supplementary Movies 3-4).

The distinct yielding modes can be quantitatively understood considering the sliding energy curves in Fig. 3b. The AB to BA and AA′ to AB′ transition energy barriers are 1.39 and 1.97 meV/atom, respectively, resulting in a ratio of ~0.71, in correspondence with the asymmetric torque ratio (Fig. 2e). Moreover, near 0° (AA′), the formed metastable state AB′ has preference to shift back to AA′ stacking (see Fig. 3b), moving opposite to the rotation. While at 60° (AB), once overcoming an intermediate (saddle point) state, the flake abruptly slips into the BA state, hence, giving rise to a sharp stick-slip behavior. Overall, the formation of intermediate metastable state of AB′ stacking acts as a strengthening effect that enhances the peak torque near 0°, leading to the pronounced torque contrast between AA′ and AB stacking states.

The above analysis indicates that elasticity plays a central role in the torsional behavior of layered interfaces. To demonstrate this, we compare the torsional response of rigid and flexible *h*-BN homojunction models. As shown in Fig. 4a, in contrast with the pronounced stick-slip behavior, characterizing the flexible interface that allows for atomic reconstruction (Fig. 3c-d), the rigid model exhibits a smooth oscillatory torque profile with comparable peak torques at AA′ and AB, accompanied by a continuous evolution of the interfacial moiré pattern. Notably, these torque peaks are 5-6 folds higher than those obtained for the flexible interface, and they exhibit different scaling with contact area ($T \propto A^{1.5}$, see SI Sec. 6).

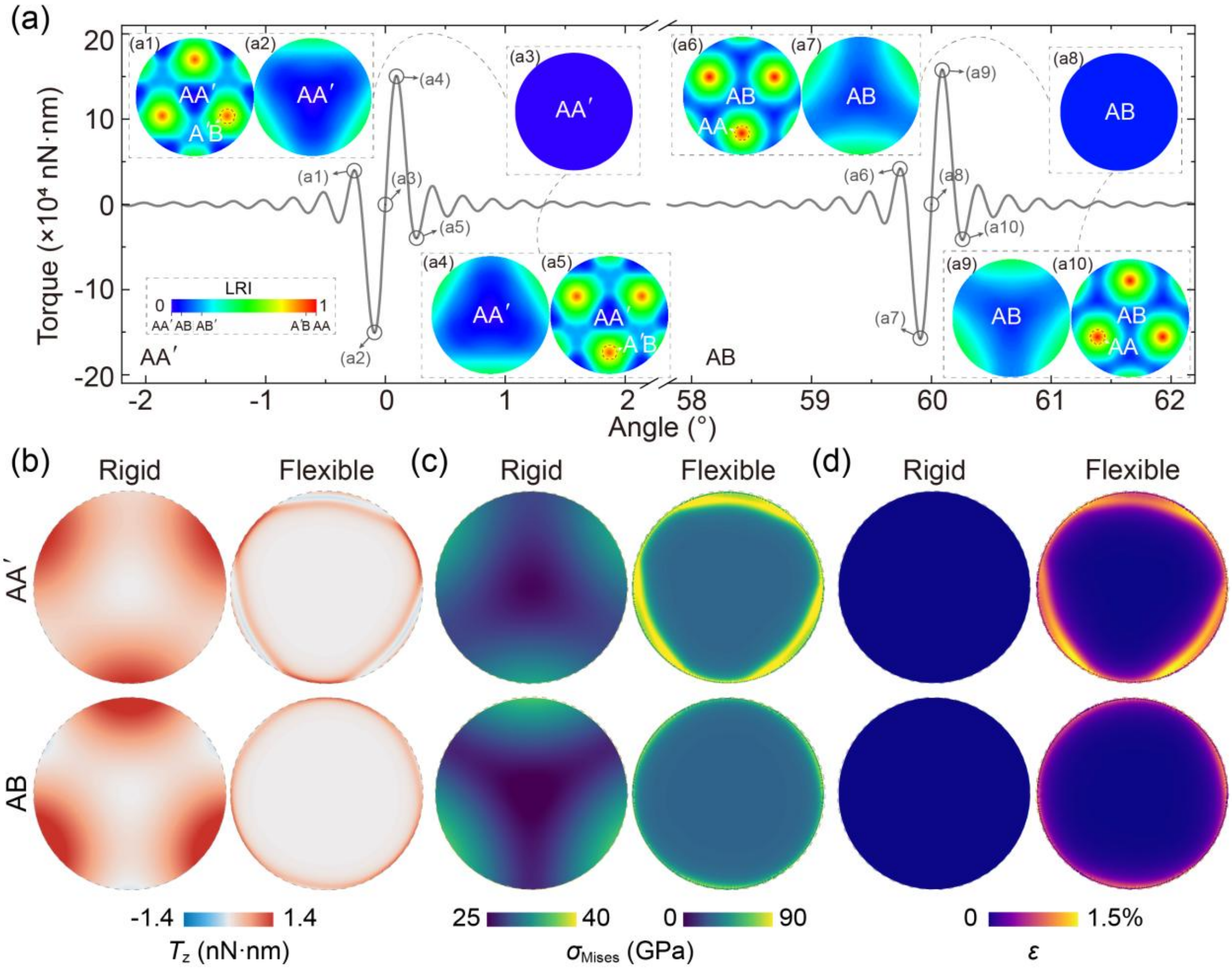

**Figure 4.** Comparison of rigid and flexible $h$-BN/$h$-BN contacts torsional responses along the rotational pathway AA′→AB→AA′. (a) Torque–angle profiles near 0° and 60° from the rigid model. The insets present the LRI maps at the marked points along the torque curves. (b)-(d) The spatial distributions of (b) local torque per atom ($T_z$), (c) von Mises stress ($\sigma_{\text{Mises}}$) and (d) in-plane deviatoric strain ($\varepsilon$) of the flake atoms at the maximum torque AA′ (top row) and AB (bottom row) configurations. For visual clarity, the torque value of each flake atom is averaged over its nearest neighbours. All simulations are performed for circular $h$-BN flakes with $R =$ 50 nm.

Figure 4b-c present the spatial distributions of local torque ($T_z$), von Mises stress ($\sigma_{\text{Mises}}$), and in-plane deviatoric strain ($\varepsilon$) at the peak torque angles. In the rigid model, the atomic torque delocalizes over the A′B and AA regions of the interface, whereas in the flexible system, it concentrates near the contact periphery. Notably, the flexible contact demonstrates a positive torque front (brown arcs) near 0° (AA′) that encase weak negative torque lines (faint blue arcs) associated with rim AB′ stacking regions that oppose the rotation. Correspondingly, Fig. 4c shows that the stress distributes more uniformly in the rigid model, whereas the AB′ regions in the flexible system manifest enhanced stress and shear strain (Fig. 4d). Notably, atomic reconstruction results in overall reduced torque values compared to the rigid case, with torque asymmetry between 0° and 60° twist angles. Moreover, the edge-dominant character of the torque distribution in the flexible case, yields a peak torque scaling of $A^{1.0}$ ($R \times R$, where one $R$ contribution comes from the explicit torque expression and another from the scaling of the rim length) rather than $A^{1.5}$ ($R^2 \times R$) for the rigid case, where the torque is delocalized over the entire contact surface.

**Concluding Remarks**

The above results demonstrate deterministic mechanical switching of twisted $h$-BN homojunctions between different commensurate states (previously shown to exhibit distinct out-of-plane electric polarization) via continuously applied torque. Unlike homonuclear graphitic interfaces that demonstrate six-fold rotational symmetry, the heteronuclear $h$-BN homojunction exhibits three-fold symmetry with peak torque asymmetry at adjacent commensurate configurations. This observation is associated with different interfacial yielding mechanisms that involve nucleation of metastable stacking states, reminiscent of brittle and ductile fracture behaviors. The maximal torque values corresponding to both commensurate states scale linearly with the contact area, highlight the dominant role of edge elasticity. This work, therefore, paves the way for the rational design of dynamically reconfigurable twistronic nanodevices demonstrating continuously tunable polarization.

**Acknowledgments**
X.G. acknowledges the financial supports from the National Natural Science Foundation of China (No. 12402133), National Key Project (No. MJZ5-2N22), and the start-up fund of University of Science and Technology of China (USTC). M.U. acknowledges the financial support of the BSF-NSF grant No. 2023614. O.H. is grateful for the generous financial support of the Heinemann Chair in Physical Chemistry and Tel Aviv University Center for Nanoscience and Nanotechnology. The computation work was performed on Tianhe new generation supercomputer at the National Supercomputer Center in Tianjin and the supercomputer at the Supercomputing Center of USTC. E.K gratefully acknowledge the Israel Science Foundation (ISF) grant 3560/25 for financial assistance and the Micro & Nano Fabrication Unit (MNFU) for the nanofabrication facilities.

# Supplementary Information for
# "Twist as a Mechanical Switch for Reconfigurable Stacking in *h*-BN"

Gautham Vijayan[1†], Zhaoheng Zhang[2†], Kenji Watanabe[3], Takashi Taniguchi[3], Michael Urbakh[4], Oded Hod[4], Xiang Gao[2*], Elad Koren[1*]

[1]*Nanoscale Electronic Materials and Devices Laboratory, Faculty of Materials Science and Engineering, Technion - Israel Institute of Technology, Haifa, 3200003, Israel.*

[2]*Department of Modern Mechanics, School of Engineering Science, University of Science and Technology of China, Hefei, Anhui 230026, China*

[3]*International Center for Materials Nanoarchitectonics, National Institute for Materials Science, Tsukuba, Japan*

[4]*Department of Physical Chemistry, School of Chemistry, The Raymond and Beverly Sackler Faculty of Exact Sciences and The Sackler Center for Computational Molecular and Materials Science, Tel Aviv University, Tel Aviv 6997801, Israel*

[†]These authors contribute equally to this work.

[*]Corresponding authors. Email: xianggao@ustc.edu.cn, eladk@technion.ac.il

In this supplemental material, we provide additional details on the following subjects:

1. Sample fabrication
2. Adhesion measurements and force calibration
3. Assignment of the twist angle
4. Relation between the applied and measured forces
5. Experiment in a different sample orientation
6. Analytical theory of torque under rigid and continuum assumption
7. MD simulation details
8. Additional high-symmetry rotation paths and their structural evolution
9. Torque–angle profiles for low-symmetry rotational paths
10. Torque–angle profiles of flakes with large radii

## 1. Sample fabrication

High-quality hexagonal boron nitride (*h*-BN) flakes were mechanically exfoliated onto prepatterned silicon wafers (8 × 8 $mm^2$). Flakes with thicknesses of approximately 250 nm were selected and marked for fabrication. PMMA resist layer was then spin-coated onto the substrate to a total thickness of ~200 nm. Electron-beam lithography was used to define arrays of circular and bearing-like structures on the selected *h*-BN flakes, with feature radii ranging from 200 to 500 nm. Following development in MIBK and IPA solution, a brief oxygen plasma treatment was applied to remove residual polymer. Metal contacts were subsequently deposited as a trilayer stack comprising 5 nm Cr, 25 nm Ni, and 25 nm Au. Lift-off in acetone removed the excess metal, yielding well-defined arrays of circular and bearing-like contacts on the *h*-BN surface. Finally, reactive ion etching was carried out using $SF_6$ (50 W, 20 sccm and 5 mTorr), with the metal contacts serving as an etch mask, resulting in an etch depth of approximately 100 nm. A schematic of the fabrication workflow is provided in Fig. S1.

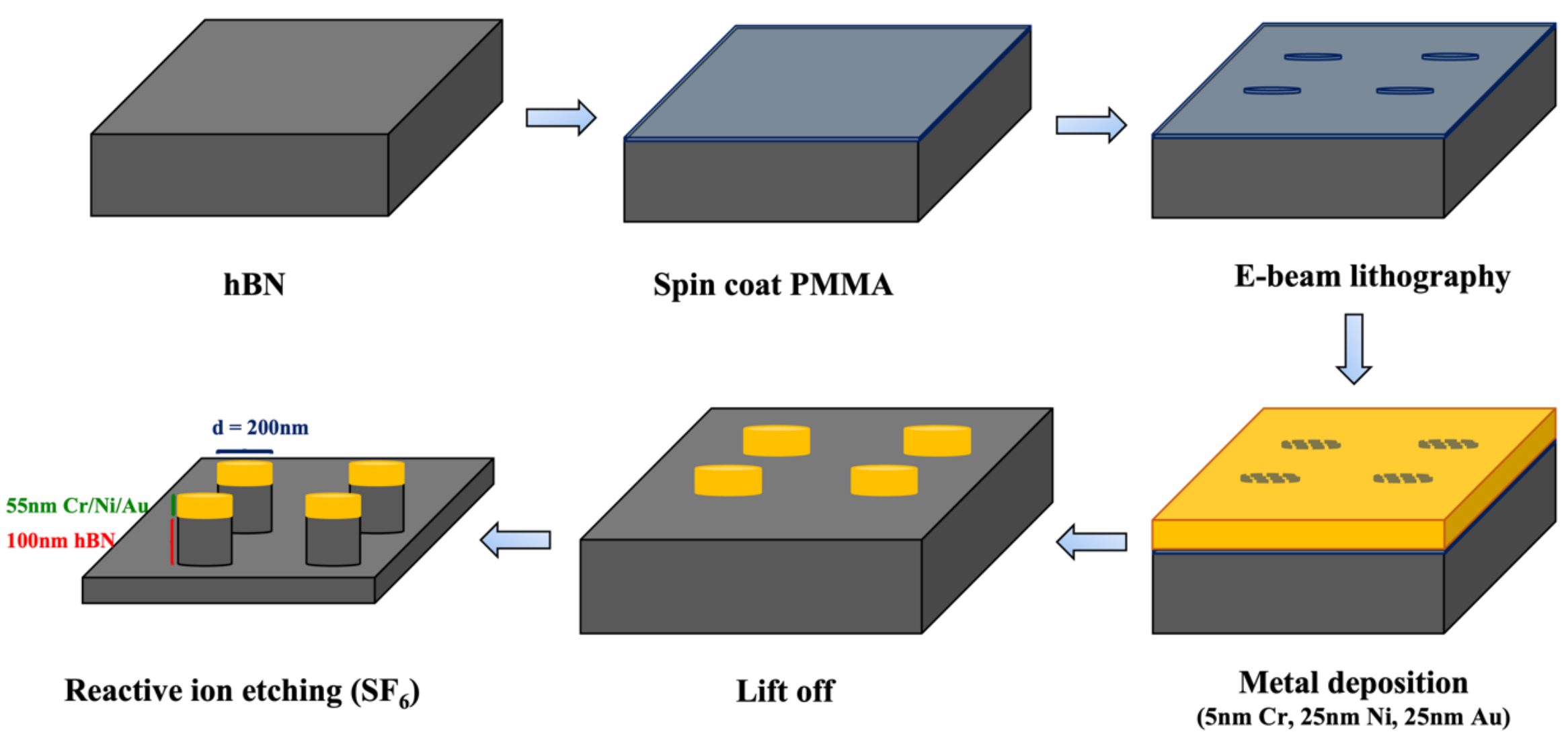

Figure S1. Schematics of the fabrication workflow.

## 2. Adhesion measurements and force calibration

A single AFM cantilever (PPP-NCLPt, Nanosensors) was used throughout all torsional and adhesion measurements. Adhesion was measured on circular $h$-BN structures of different radii by laterally shearing the pillars with the AFM tip (inset in Fig. S2a) while recording force-distance curves (Fig. S2a). Upon contact, the lateral force increases with cantilever deflection; further lateral displacement shears the top $h$-BN layer relative to the bottom layer, creating a pristine $h$-BN/$h$-BN interface. At the onset of sliding, the force drops abruptly and then decreases further as the overlap area is reduced. The lateral force at sliding onset, corresponding to maximum overlap, is given by $F = 2\sigma R$, where $R$ is the interface radius and $\sigma$ is the adhesion energy[1–4]. Using the previously established linear dependence of lateral force on contact radius[1], we measured circular $h$-BN pillars with radii of 200, 300, and 400 nm. A linear fit of the measured lateral-deflection signal (in volts) as a function of radius (Fig. S2b) yielded $V/R = 3.3277\times10^{-4}$ V·nm$^{-1}$. Using the known adhesion energy of $h$-BN ($\sigma = 0.372$ J·m$^{-2}$)[5], a corresponding force-radius relation is obtained as $F/R = 2\sigma = 0.744$ nN·nm$^{-1}$. The lateral-force calibration ratio ($C$) was therefore obtained by comparing the expected force-radius slope with the experimentally measured voltage-radius slope as, $C = F/V = (F/R)/(V/R) = 2.236\times10^{3}$ nN·V$^{-1}$. This calibration ratio was then used for all subsequent torque measurements acquired with the same cantilever.

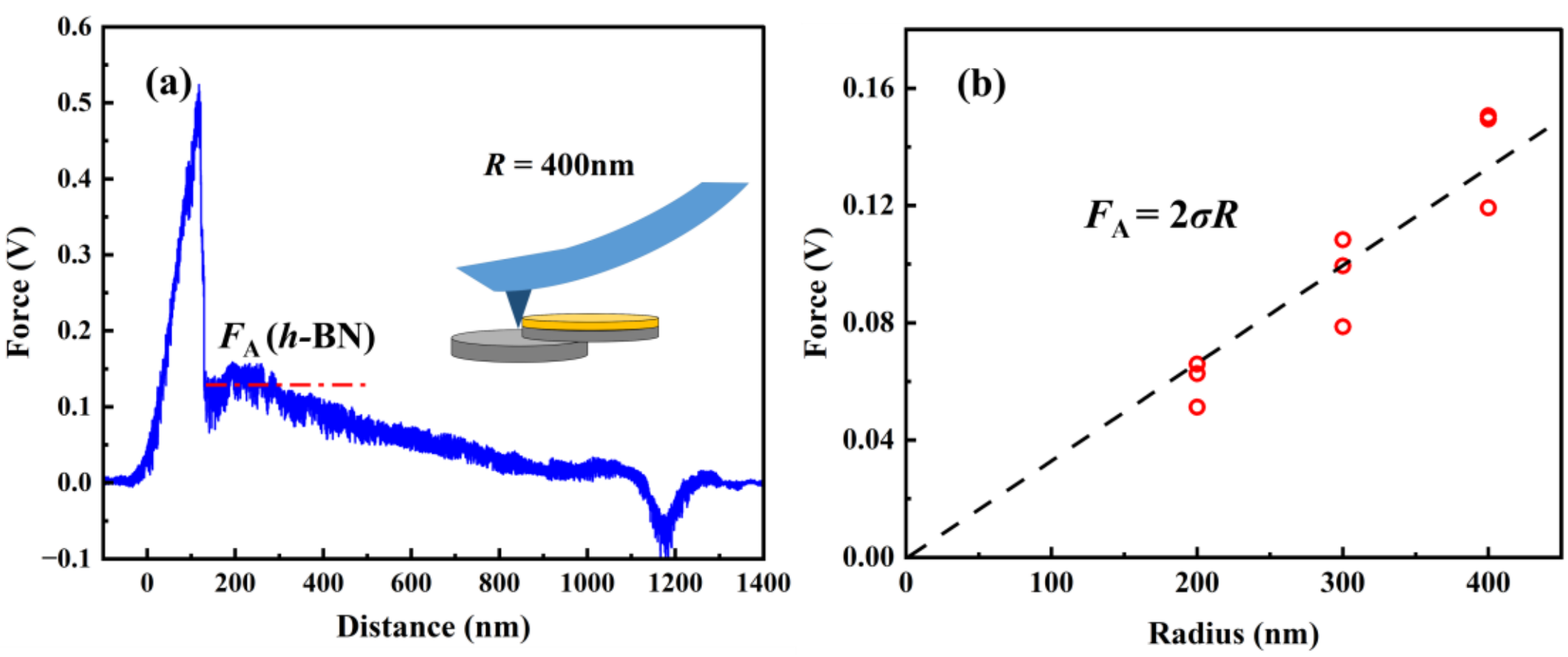

Figure S2. (a) A representative force-distance curve recorded during an adhesion measurement. The peak lateral force at the onset of shearing was used to extract the adhesion energy. The inset shows a schematic of the shearing process. (b) Scaling of maximal lateral force with contact radius. The black dashed line represents a linear fit with a slope of $3.3277\times10^{-4}$ V·nm$^{-1}$.

## 3. Assignment of the twist angle

During the experiment, the oscilloscope records the lateral force as a function of time. To convert the time-dependent signal into angular displacement, the Y-axis piezo displacement of the cantilever was recorded simultaneously, as shown in Fig. S3. Since the Y-axis displacement changes only during the circular actuation, the time interval $\Delta t$ between the starting point $(x_1, y_1)$ and the end point $(x_2, y_2)$ corresponds to the time duration required to traverse the predefined angular arc $\Delta\theta$. The instantaneous angular position associated with the lateral force trace can therefore be expressed as $\theta(t) = t{\times}\Delta\theta/\Delta t$, where $t$ is the elapsed time measured from the onset of circular motion and we assume constant angular velocity.

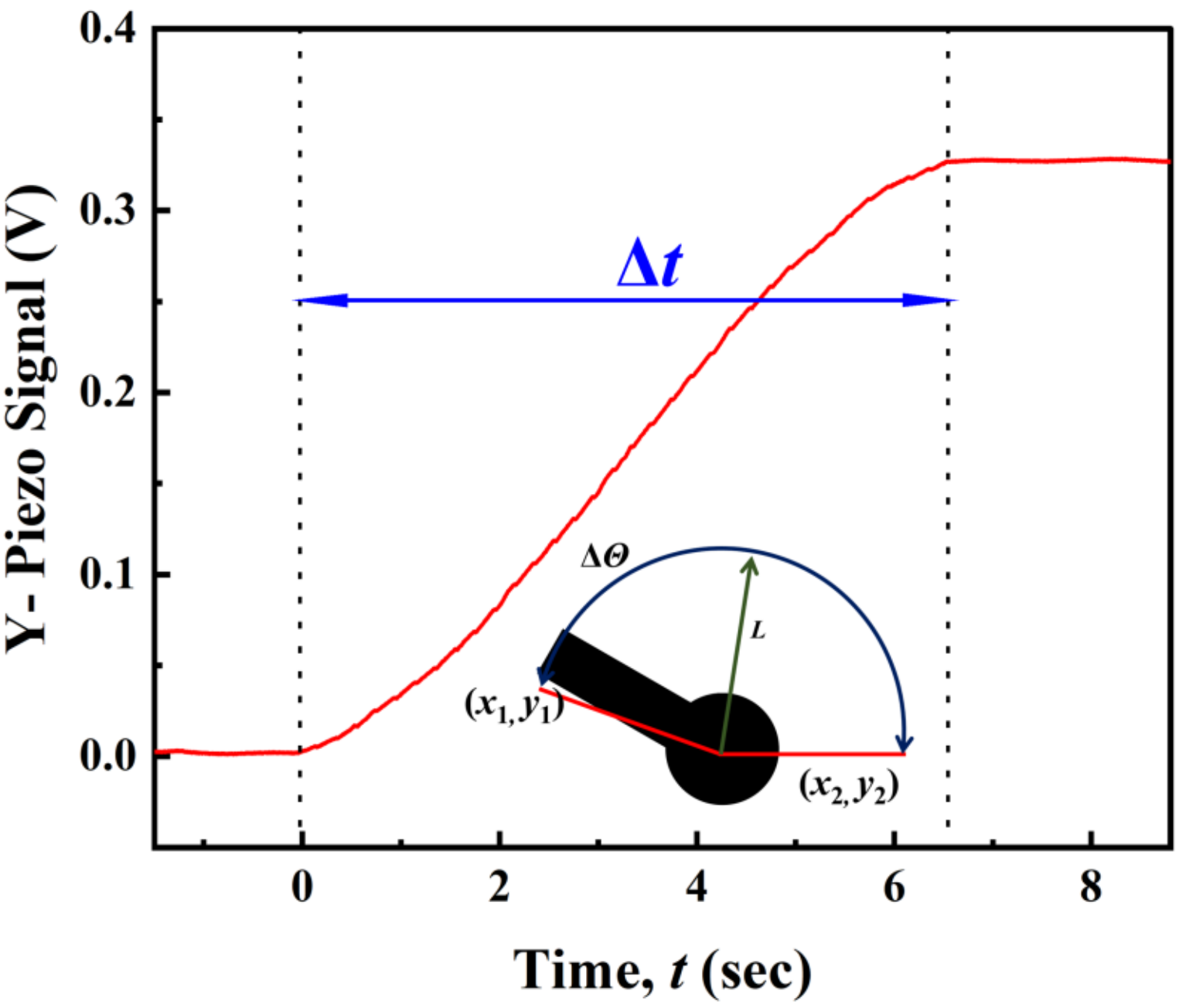

Figure S3. Oscilloscope trace of cantilever displacement along the Y axis during actuation. The blue arrow marks the time required to complete the actuation cycle.

## 4. Relation between the applied and measured forces

The sample orientation was chosen such that, at the commensurate configurations (60° and 120°), the lateral force component ($F_x$) measured by the cantilever is related to the applied force via $F_x = F_{\text{real}} \times \cos(30°)$ for both commensurate angles, as shown in Fig. S4a. In other words, the angle between the measured lateral component and the resultant force is identical at both commensurate orientations. A cosine correction was therefore applied to recover the true force at each commensurate position. The corresponding torque was then determined as the product of this corrected force and the radius of the circular trajectory.

The force profile over the full (360°) range is presented in Fig. S4b, from which the force profile in Fig. 1c of the main text was extracted. In principle, the tangential forces at 0°, 180°, and 360° are expected to be zero. Nonetheless, the small peak observed at 0° results from the force required to break the pillar and initiate rotation.

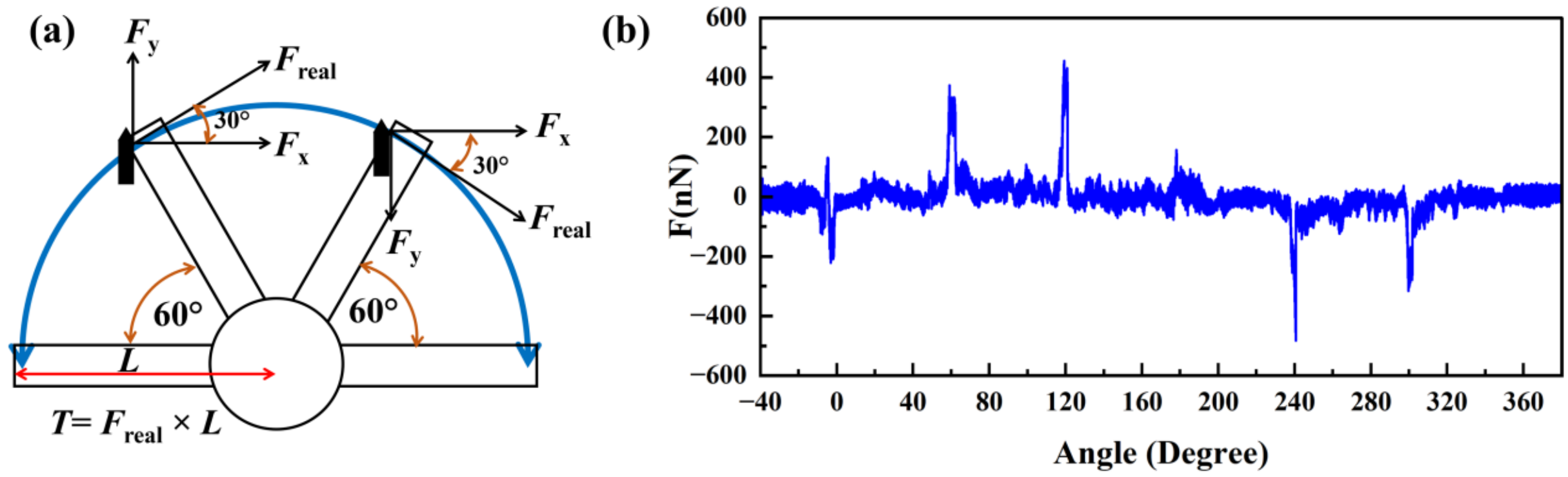

Figure S4. (a) Geometric relation between the force applied by the tip on the cantilever and the measured component, $F_x$, during circular motion. (b) A full (360° range) force-angle curve including the initial breaking stage for a pillar of circular radius of 500 nm.

## 5. Experiment in a different sample orientation

To validate that the measured response reflects the evolving interaction between the cantilever and the sample, force-angle profiles were acquired for an alternative sample orientation, as shown in Fig. S5a. In this geometry, the commensurate positions are labelled 2 and 3, corresponding to 60° and 120°, respectively. The cantilever-sample interaction and the associated force analysis for this configuration, illustrated in Fig. S5b, differ from those shown for the previous orientation (Fig. S4). The measured force-angle profile for an interfacial radius of 300 nm is presented as the red curve in Fig. S5c, and the corresponding AFM topography obtained after rotation is shown in Fig. S5d. Notably, the two force peaks in the measured profile are of nearly equal magnitude, in contrast to the behaviour observed for the earlier sample orientation. However, after applying the required geometric correction, the force-angle profile becomes the same as that obtained for the abovementioned orientation, exhibiting a smaller peak at 60° followed by a larger peak at 120°. This confirms that the force-angle profile is a signature of the intrinsic interfacial registry, independent of sample orientation and geometry.

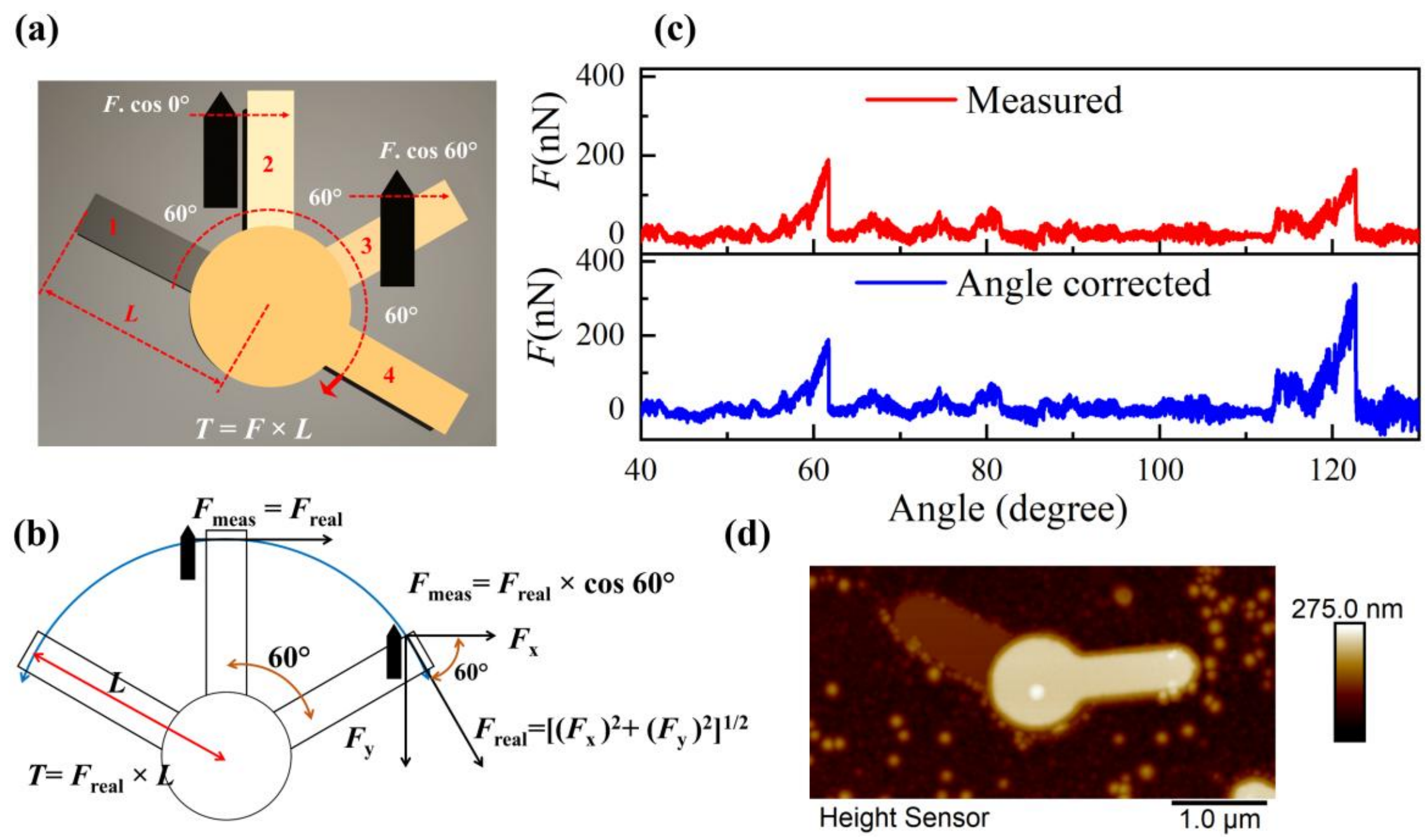

Figure S5. (a) Schematic illustration of the sample orientation and cantilever-sample interaction geometry. (b) Corresponding force analysis of the cantilever-sample interaction. (c) Measured (red) and corrected (blue) force-angle profiles. (d) AFM topography of the structure after rotation.

## 6. Analytical theory of torque under rigid and continuum assumption

The periodic moiré-level interlayer potential energy per unit area at twist angle $\theta$ can be expressed as[6,7]:

$$W(\boldsymbol{X};\theta) = U_{\mathrm{ref}} + \Re\left(\sum_{n=1}^{3} U e^{i\boldsymbol{G}_n(\theta)\cdot\boldsymbol{X}}\right), \tag{1}$$

where $\boldsymbol{X}$ is the spatial coordinate vector within the moiré superlattice, $U_{\mathrm{ref}}$ is the reference energy density, $U = U_0 - iU_1$ (where $U_0$ and $U_1$ are real) represents the complex Fourier coefficient, and $\Re(\cdot)$ represents the real part of a complex number. The reciprocal moiré lattice vectors, $\boldsymbol{G}_n(\theta)$, are given by:

$$\boldsymbol{G}_n(\theta) = \frac{4\pi}{\sqrt{3}\lambda(\theta)}\boldsymbol{\mathcal{R}}(\psi)\hat{\boldsymbol{x}}, \tag{2}$$

where $\lambda(\theta) \approx a_{\mathrm{BN}}/\left[2\sin\left(\frac{\theta}{2}\right)\right]$ is the moiré period, $a_{\mathrm{BN}}$ is the lattice period of $h$-BN, $\theta$ is the twist angle of the flake, $\boldsymbol{\mathcal{R}}(\cdot)$ denotes the rotation matrix, $\hat{\boldsymbol{x}}$ is the unit vector along the $x$-axis, and the angle $\psi = \frac{\theta}{2} + \frac{\pi}{2}$ characterizes the rotation of the moiré superlattice. For a flake with its center-of-mass (C.O.M) located at $\boldsymbol{x}_0$, the total interlayer potential energy $E$ is the area integral of the energy density function over the occupied domain $\Omega$:

$$E = \int_{\Omega(\boldsymbol{x})} W(\boldsymbol{X};\theta)\mathrm{d}A. \tag{3}$$

For a circular flake, this translates to:

$$E_{\mathrm{cir}} = \pi R^2 U_{\mathrm{ref}} + \frac{\sqrt{3}}{2}\lambda R J_1\left(\frac{4\pi R}{\sqrt{3}\lambda(\theta)}\right)\Re\left[\sum_{n=1}^{3} U e^{i\boldsymbol{g}_n\cdot\boldsymbol{x}_0}\right], \tag{4}$$

where $J_1(\cdot)$ is the Bessel function of the first kind, and $\boldsymbol{g}_n = \frac{4\pi}{\sqrt{3}a_{\mathrm{BN}}}\boldsymbol{\mathcal{R}}\left(\frac{2(n-1)\pi}{3}\right)\hat{\boldsymbol{x}}$ represents the primitive reciprocal lattice vectors of the $h$-BN monolayer. The external torque, $T_{\mathrm{cir}}$, required to overcome that induced by interlayer potential, equals to the positive derivative of the total interlayer potential energy with respect to the twist angle,

$$T_{\mathrm{cir}} = +\frac{\partial E_{\mathrm{cir}}}{\partial\theta} = -\pi R^2\cot\left(\frac{\theta}{2}\right)J_2\left(\frac{4\pi R}{\sqrt{3}\lambda(\theta)}\right)\Re\left[\sum_{n=1}^{3} U e^{i\boldsymbol{g}_n\cdot\boldsymbol{x}_0}\right]. \tag{5}$$

By applying a small-angle approximation ($\theta \ll 1°$), we extract the peak torque $T_{\mathrm{cir}}^{\max}$ to establish the explicit area ($A$) scaling law as:

$$T_{\mathrm{cir}}^{\max}(A) = \frac{1.47}{a_{\mathrm{BN}}}A^{1.5}\left|\Re\left[\sum_{n=1}^{3} U e^{i\boldsymbol{g}_n\cdot\boldsymbol{x}_0}\right]\right|. \tag{6}$$

To parameterize the analytical model, the energy amplitudes, $U_0$ and $U_1$, were calibrated using the registry-dependent ILP for the interlayer vdW interactions[8–11]. By evaluating the energy differences among the high-symmetry stacking states at a fixed rigid interlayer distance of $d = 3.35$ Å, these energy amplitudes were extracted as $U_0 = 1.63$ meV/Å$^2$ and $U_1 = 0$ meV/Å2 for aligned $h$-BN/$h$-BN homostructure, whereas $U_0 = 1.21$ meV/Å$^2$ and $U_1 = -0.20$ meV/Å$^2$ for anti-aligned configurations.

By using these parameters, the resulting analytical energy and torque landscapes, as well as the predicted scaling results of $T_{\text{cir}}^{\max}$ agree well with the corresponding rigid atomistic calculations (Fig. S6). The corresponding scaling pre-factors follow the order AB′ < AA′ ≃ AB < A′B < AA.

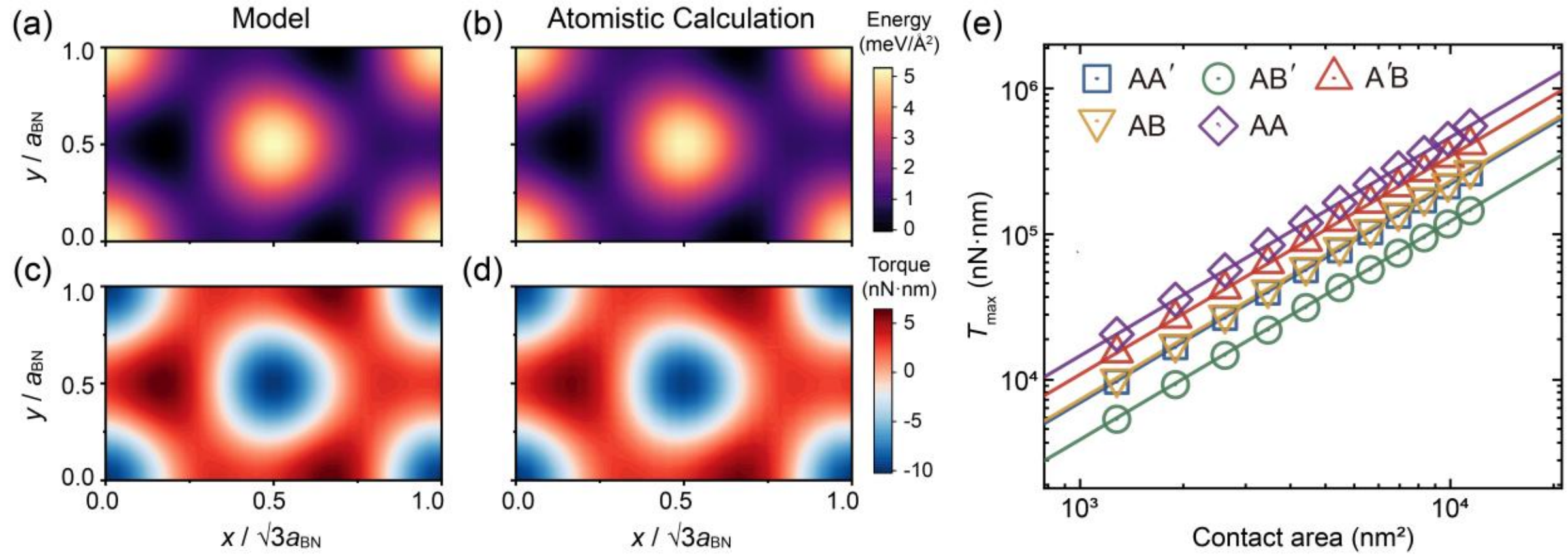

Figure S6. Comparison of analytical model and rigid atomistic calculation results. (a)-(b) The relative interfacial energy density for a circular anti-aligned $h$-BN/$h$-BN flake ($R = 20$ nm) with a twist angle of $\theta = 0.1°$, obtained from the (a) Eq. S4 and (b) MD, respectively. (c)-(d) Corresponding torque distributions from (c) Eq. S5 and (d) MD, respectively. The interlayer separation was fixed at 3.35 Å. (e) Scaling of peak torque, $T_{\max}$, with contact area, for circular flakes of different stacking states. Data points are obtained from rigid atomistic calculations. Solid lines are predictions from the analytical rigid model (Eq. S6), both demonstrating a scaling of $T_{\max} \propto A^{1.5}$. The interlayer separation was fixed at 3.35 Å.

## 7. MD simulation details

All classical molecular dynamics (MD) simulations were performed using the LAMMPS package[12]. The flexible MD model comprises an $h$-BN substrate, a finite-sized circular $h$-BN flake, and a rigid driving ring (1 nm in width) replicated from the flake to apply rotational displacement. For simulating different rotation paths, slightly different flakes are cut out with geometric centers located exactly at the target rotation centers. Periodic boundary conditions were applied in the lateral dimensions. The intra- and interlayer interactions of the flake and the substrate were described using the Tersoff[13] potential and anisotropic interlayer potential (ILP)[8–11], respectively. Torsional actuation was introduced by harmonic springs that connect the rigid driving ring to the underlying flake atoms with in-plane spring stiffness $k_{xy}^{d} = 0.015$ eV/Å$^2$, calibrated to approximate the interlayer shear modulus (~3.0 GPa) of graphene layers (Fig. S7a-S7b). To mimic the effect of multilayer stacks, harmonic constraints ($k_z = 0.167$ eV/Å$^2$, Fig. S7c-S7d) that approximate the normal modulus (~33.4 GPa) of graphene layers were applied to the vertical degrees of freedom of all the flake atoms. Recently, we calculated the corresponding effective spring stiffnesses for $h$-BN (unpublished results) finding them to be very close to those used herein with minor effects on the extracted yielding strengths. Furthermore, to account for the elastic compliance of the substrate, its atoms were constrained to their equilibrium positions via the same harmonic springs (i.e., $k_{xy} = 0.015$ eV/Å$^2$ and $k_z = 0.167$ eV/Å$^2$).

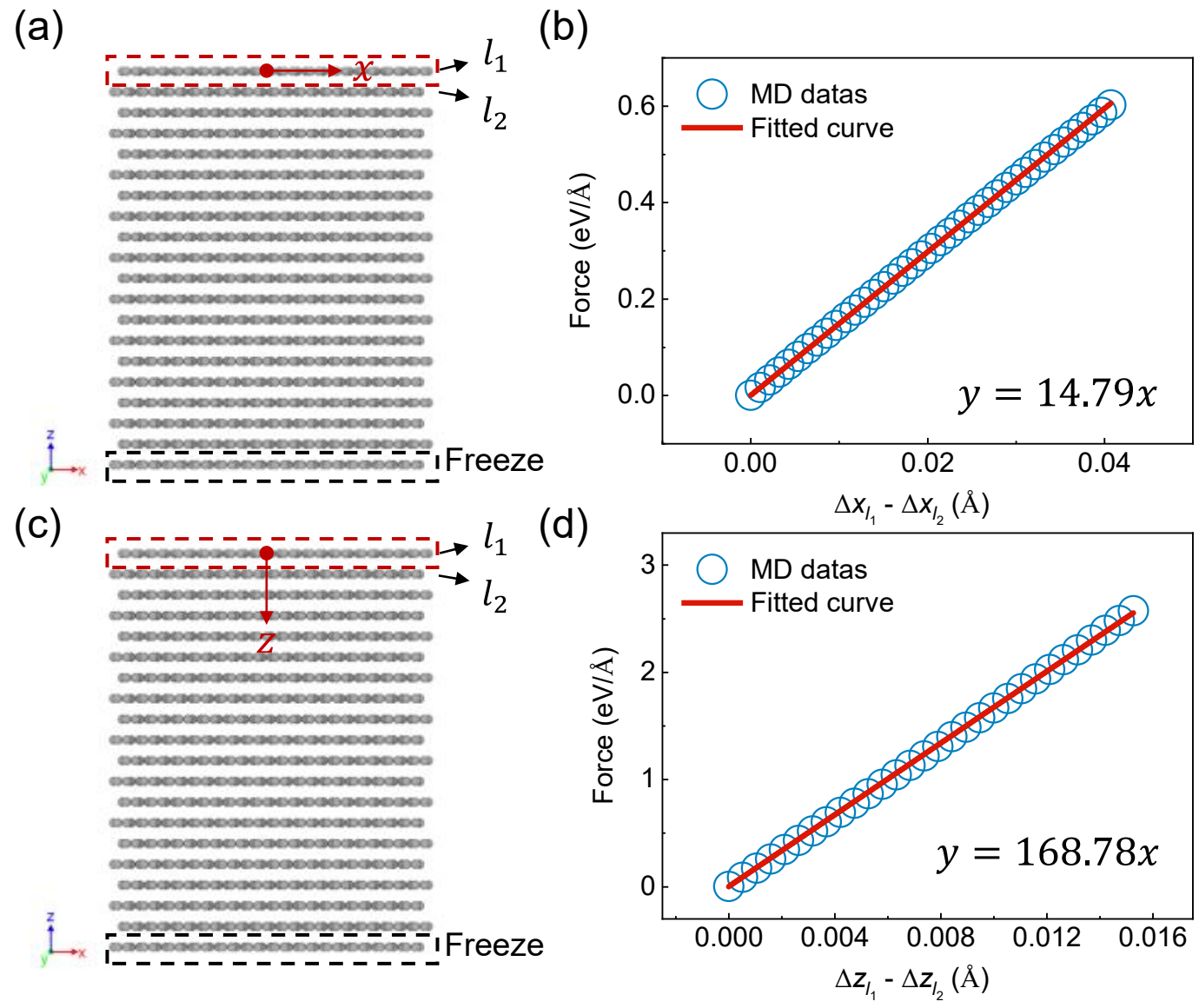

Figure S7. Calibration of the effective harmonic spring constants by using a periodic 5 nm × 5 nm, 20-layer Bernal-stacked graphene model with an interlayer spacing of 3.37 Å. (a)-(b) In-plane shear along the armchair direction and the corresponding force-displacement curve. (c)-(d) Out-of-plane loading and the corresponding force-displacement curve. Per-atom harmonic spring stiffness is calculated as the total stiffness divided by the total number of atoms (1008) in each layer.

Quasi-static torsional simulations were performed using a multistep energy minimization protocol. Following a full relaxation of initial state, the rigid driving ring was incrementally rotated in angular steps, such that the arc displacement of the outer rim atoms does not exceed 0.1 Å, regardless of the flake radius. After each rotational increment, the coordinates of the driving ring were fixed, and the remainder of the system was allowed to fully relax until energy convergence was achieved. Each relaxation stage employed an hybrid optimization strategy with an initial coarse minimization using the conjugate gradient algorithm with a stringent tolerance of $1.0\times10^{-20}$ (eV or eV/Å), followed by a refined minimization utilizing the FIRE algorithm[14] with a force convergence tolerance of $1.0\times10^{-3}$ eV/Å. Thereafter, the torque of the driving ring was calculated relative to the rotation center.

For comparative analysis, rigid atomistic calculations, consisting solely of a rigid *h*-BN flake and a rigid *h*-BN substrate (without a driving ring), were also performed. In the rigid calculations, the flake was rotated with a constant angular step of 0.01° at a fixed rotation center and a constant interlayer distance of 3.35 Å, without any structural relaxation or lateral motion.

Per-atom von Mises stress ($\sigma_{\mathrm{Mises}}$) and in-plane deviatoric strain ($\varepsilon$) were calculated to characterize the local mechanical response of the flake. The stress was evaluated from the per-atom virial stress tensor after removing the contribution of the springs connecting the driving ring to the rotating flake, so that the stress field reflects the intrinsic deformation of the flake due to the interlayer interactions with the substrate. For each atom, the von Mises stress was calculated as

$$\sigma_{\mathrm{Mises}} = \sqrt{\frac{1}{2}\left[\left(\sigma_{xx}-\sigma_{yy}\right)^2 + \left(\sigma_{yy}-\sigma_{zz}\right)^2 + \left(\sigma_{zz}-\sigma_{xx}\right)^2 + 3\left(\sigma_{xy}^2+\sigma_{xz}^2+\sigma_{yz}^2\right)\right]}. \quad 7)$$

For better visualization, the atomic von Mises stress was further smoothed by a nearest-neighbor weighted average, with the central atom and neighboring atoms assigned weights of 1 and 1/3, respectively.

The in-plane strain field was obtained using the two-dimensional atomic strain analysis in OVITO package[15], where the local deformation gradient was fitted relative to the relaxed initial configuration. The in-plane deviatoric strain was then defined from the fitted strain components as

$$\varepsilon = \sqrt{\varepsilon_{xy}^2 + \frac{1}{2}\left(\varepsilon_{xx}-\varepsilon_{yy}\right)^2}, \quad 8)$$

which reflects the local shear strain intensity.

## 8. Additional high-symmetry rotation paths and their structural evolution

To further validate the proposed rotational path (AA′→AB→AA′), we evaluate the torque-angle profiles across various rotational pathways initiated from the stable AA′ and AB stacking states. To emphasize the effect of elasticity, results obtained considering rigid flake rotation on rigid substrate (without driving ring) are also provided. To preserve the threefold rotational symmetry of the finite circular contact, the rotation axes need to pass through high symmetry stacking sites and hollow centers, as shown in Fig. S8a and S8f. Aside from P1: AA′→AB→AA′ (Fig. S8b and S8d), another rotational path initiated at AA′ (i.e., P2: AA′→AA→AA′) yields asymmetry under rigid twist. However, the peak torque at 0° is half of that at 60° (Fig. S8c), opposite to the experimental observations. Moreover, this asymmetric rotational profile becomes symmetric when elastic relaxation is allowed (Fig. S8e). Because the AA stacking configuration is energetically the least stable, this path can be safely excluded.

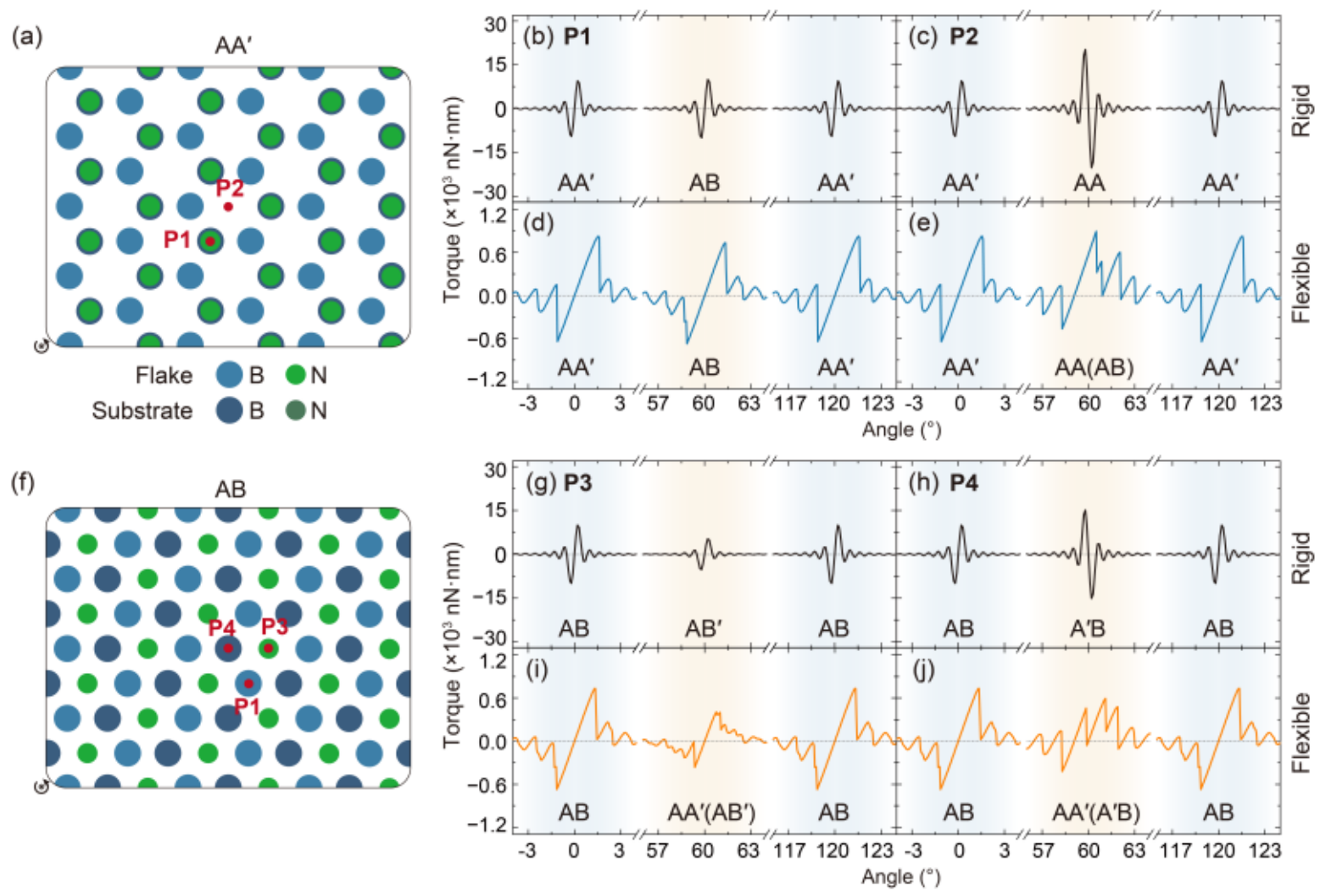

Figure S8. Torsional profiles across different rotational pathways and initial stacking states. (a) Schematic of the h-BN flake in the initial AA′ stacking, where blue and green speres denote B and N atom, respectively, with lighter and darker colors representing atoms in the upper flake and lower substrate layers, respectively. The red dots denote the rotational centers for Path 1 and Path 2 (P1 and P2), respectively. (b)-(e) Calculated torque-angle profiles starting from the AA′ stacking. (b)-(c) rigid twisting around P1 and P2, respectively. (d)-(e) flexible quasi-static twisting around P1 and P2, respectively. (f) Schematic of the flake in the initial AB stacking, highlighting three distinct rotation centers (P1, P3, and P4), using the same color scheme as in panel a. (g)-(j) Same as (b)-(e) but for torque-angle profiles twisting around P3 and P4 starting from the AB stacking mode. All simulations/calculations are performed on circular h-BN flakes with $R = 10$ nm.

Asymmetric periodic torque also emerges in pathways originating from the AB stacking state, such as Path 3 [rigid: AB→AB′→AB, flexible: AB→AA′(AB′)→AB] and Path 4 [rigid: AB→A′B→AB; flexible: AB→AA′(A′B)→AB], as shown in Fig. S8g-S8j. Although Path 3 involves the metastable AB′ state as the rotation center, which is more energetically favourable than the unstable state A′B stacking, the asymmetry between the torque peaks at 60° [AA′(AB′)] and 0° (AB) reduces with contact area (Fig. S9), in contradiction to experiments. Therefore, the rotational paths initiated from AB stacking at 0° (P3 and P4) can be also ruled out.

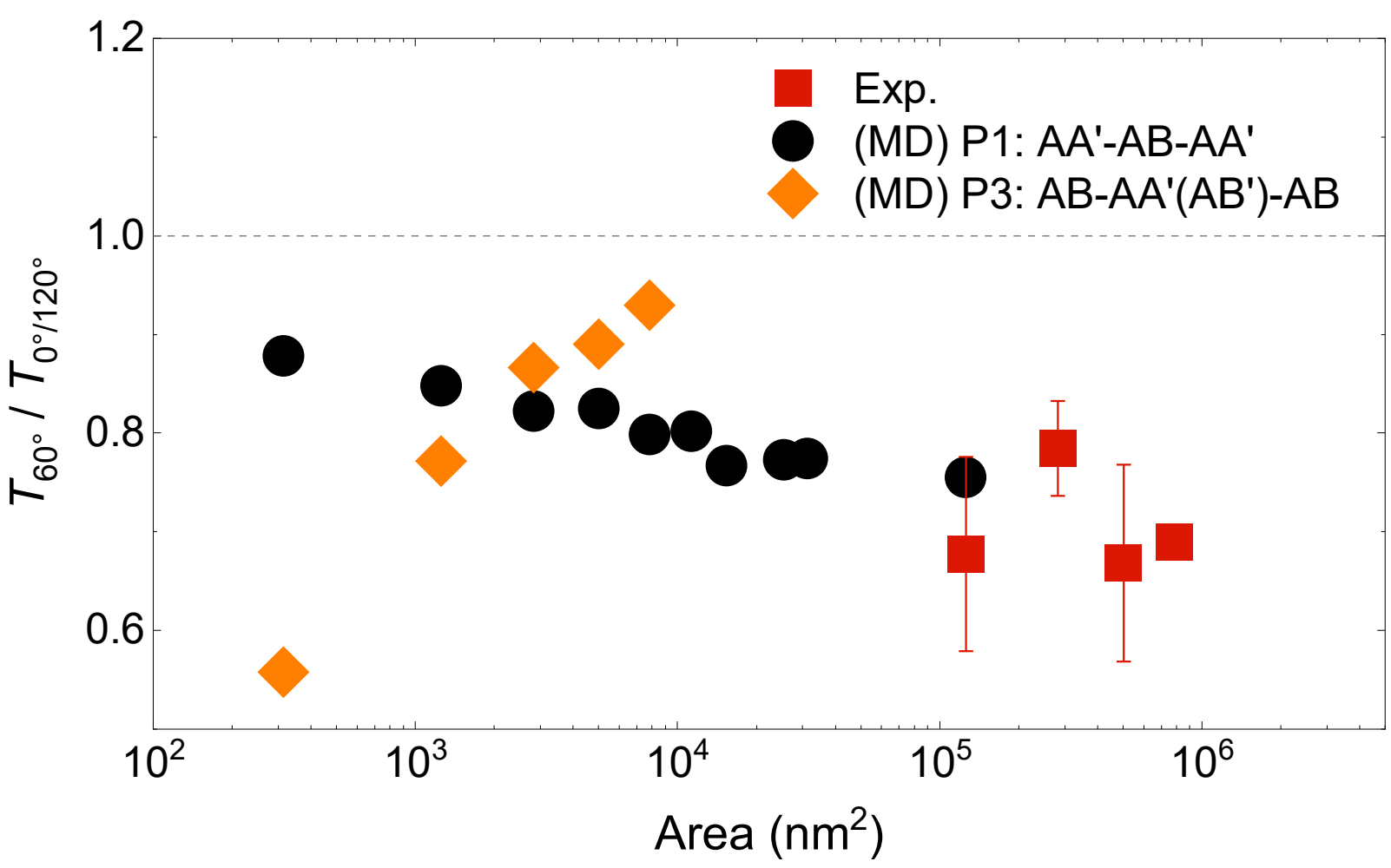

Figure S9. Comparison of peak torque ratios measured in experiments (red squares) and calculated via MD simulation for rotation paths P1: AA′→AB→AA′ (black circles) and P3: AB→AA′(AB′)→AB (orange diamonds).

To understand the structural evolution, Figure S10 presents zoomed-in torque–angle profiles with representative LRI distributions for Paths 2-4 in Fig. S8. Under atomic relaxation, the flakes stay at their AA′ and AB states near 0° (Fig. S10a, S10c, S10e), whereas near 60° they slip from these stackings to the AB and AA′ states, respectively for Paths 2-4 (Fig. S10b, S10d, S10f). The rotation paths are thus denoted as P2: AA′→AB(AA)→AA′, P3: AB→AA′(AB′)→AB, and AB→AA′(A′B)→AB, respectively, where the parentheses represent the stacking mode before and after lateral slip.

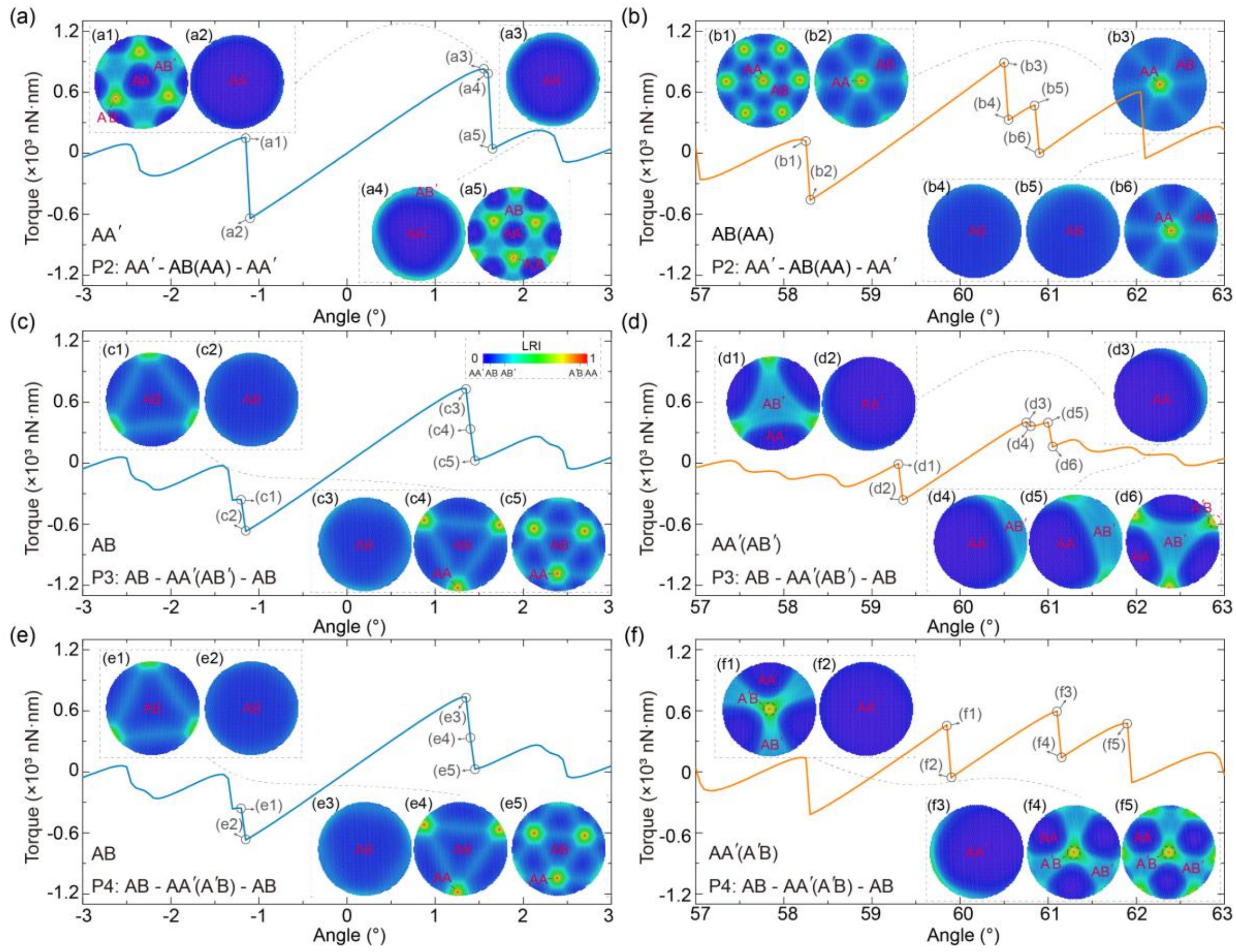

Figure S10. Structural evolution of flexible $h$-BN/$h$-BN interfaces along different rotational pathways of the $R = 10$ nm circular contact. (a)-(f) Torque–angle profiles for different initial stacking configurations near 0° (a, c, e) and 60° (b, d, f), corresponding to rotation paths P2 (a, b), P3 (c, d), and P4 (e, f) in Fig. S8. The inset maps show the corresponding LRI distributions at the marked angular positions.

## 9. Torque–angle profiles for low-symmetry rotational paths

To examine the influence of the position of the rotation-center, two additional low-symmetry centers (marked as SP1 and SP2) were selected for the AA′- and AB-stacked interfaces, respectively, as shown in Fig. S11a and S11c. As presented in Fig. S11b and S11d, the torque–angle peaks near twist angles of 0° and 120° differ quantitatively, thus suggesting that the threefold rotation symmetry observed in the experiments is violated for these off-symmetry rotation paths.

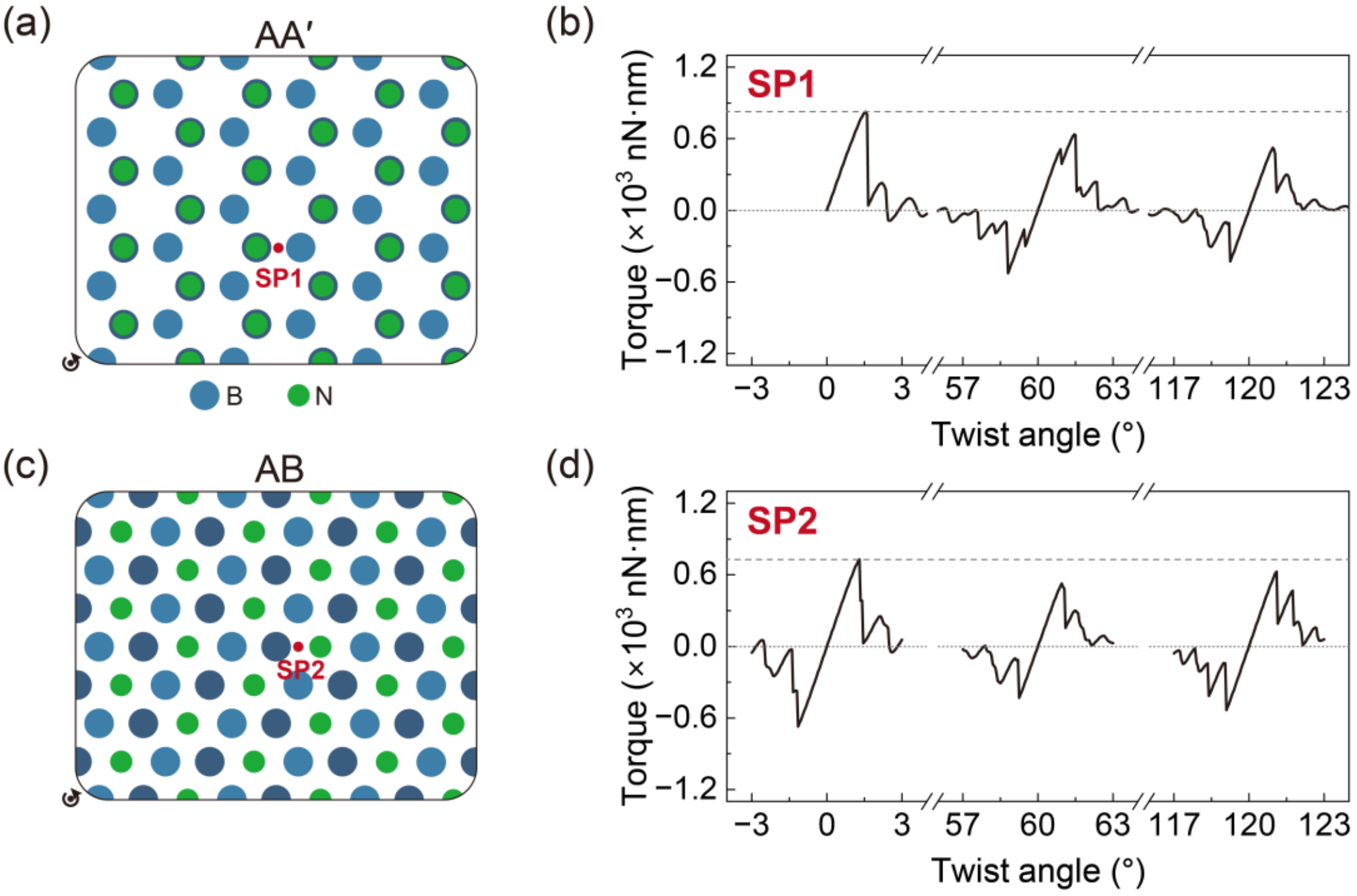

Figure S11. Rotational profiles of the $R = 10$ nm circular $h$-BN flake twisting around low-symmetry rotation centers. (a) and (c) Schematics of the flake in the initial AA′- and AB stacking modes marking the low-symmetry rotation centers SP1 and SP2. (b) and (d) The corresponding torque–angle profiles.

## 10. Torque–angle profiles of flakes with large radii

To further reveal the different yielding behaviors of the two commensurate stacking states along the AA′→AB→AA′ rotational path, Fig. S12a-S12c present additional torque–angle curves for circular flakes with large radii of 70, 100, and 200 nm obtained from MD simulations. For all three sizes, the AB state near 60° exhibits a linear-elastic-like torque–angle response followed by an abrupt instability at the peak torque, resembling a brittle fracture yielding mode. In contrast, the AA′ state near 0° deviates remarkably from the initial linear regime before final slip. Owing to the higher transition energy barrier between stable AA′ and metastable AB′ compared to that between equally stable AB and BA, the torque grows further in the linear regime (see Fig. S12) and then yields with a reduced slope (due to the formation of AB′ stacking at edges, as shown in Fig. 3c), reaches its maximum, and decreases slightly prior to sudden slip, analogous to a ductile fracture response.

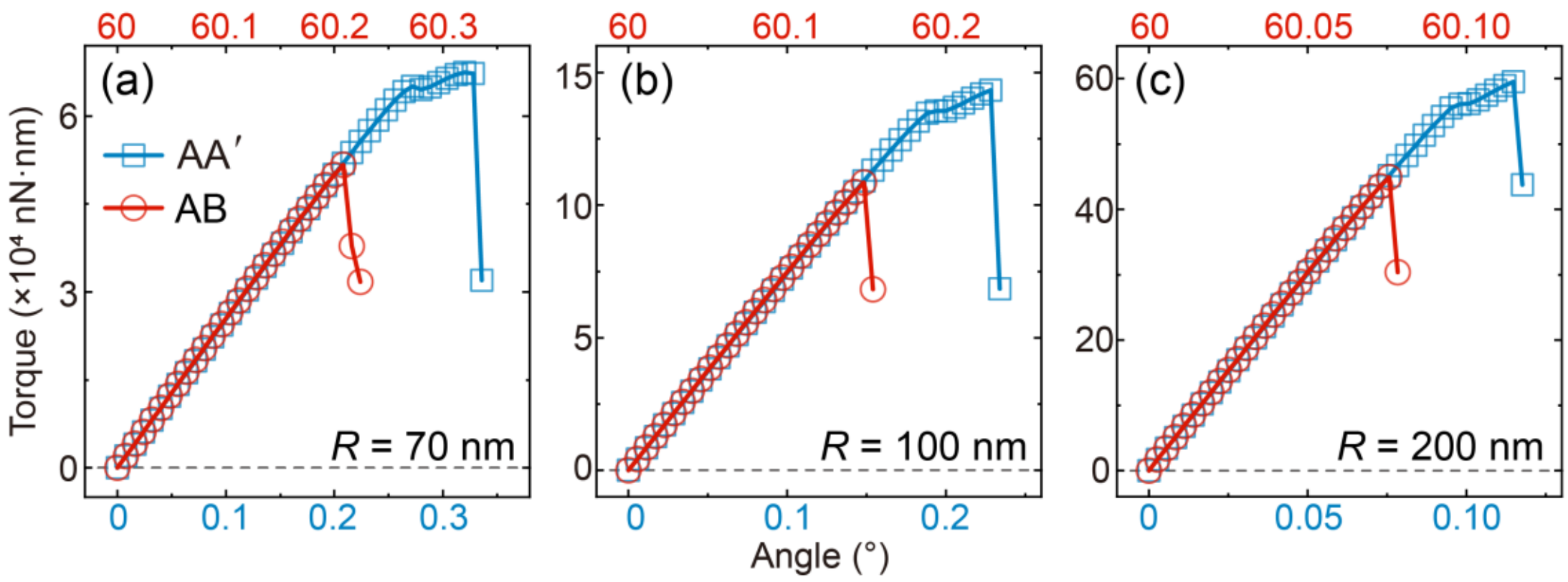

Figure S12. Torque–angle curves for circular flakes with large radii. Results are shown for (a) $R = 70$ nm, (b) $R = 100$ nm, and (c) $R = 200$ nm. The blue-square (bottom $x$-axis) and red-circle (top $x$-axis) curves represent the torque variation in the vicinity of 0°/120°(AA′) and 60°(AB) twist angles, respectively, highlighting their distinct yielding behaviors along the AA′→AB→AA′ rotational path.